\documentclass[twocolumn,showpacs,preprintnumbers,amsmath,amssymb]{revtex4}

\usepackage{graphicx}
\usepackage{dcolumn}
\usepackage{bm}

\begin{document}
\title{\large Non-invasive detection of molecular bonds in quantum dots}

\author{M. C. Rogge}
\email{rogge@nano.uni-hannover.de}
\author{R.~J. Haug}
\affiliation{Institut f\"ur Festk\"orperphysik, Leibniz
Universit\"at Hannover, Appelstr. 2, 30167 Hannover, Germany}

\date{\today}

\begin{abstract}
We performed charge detection on a lateral triple quantum dot with
star-like geometry. The setup allows us to interpret the results
in terms of two double dots with one common dot. One double dot
features weak tunnel coupling and can be understood with atom-like
electronic states, the other one is strongly coupled forming
molecule-like states. In nonlinear measurements we identified
patterns that can be analyzed in terms of the symmetry of
tunneling rates. Those patterns strongly depend on the strength of
interdot tunnel coupling and are completely different for atomic-
or molecule-like coupled quantum dots allowing the non-invasive
detection of molecular bonds.
\end{abstract}

\maketitle

Quantum dots are often called artificial atoms
\cite{Kouwenhoven-97} due to their discrete electronic level
spectrum. When several quantum dots are connected, they start to
interact \cite{Wiel-03}. If the tunneling rate between the dots is
small, the electronic wavefunctions are still constricted to the
single quantum dots and the interaction is dominated by
electrostatics with sequential interdot tunneling. In contrast,
for large tunneling rates electronic states can be found extended
over several dots. These extended states introduce covalent
bonding as in real molecules \cite{Blick-96}. Whether or not the
interdot tunneling rates are sufficient to form coherent
molecule-like states is a crucial information in order to properly
describe a quantum dot system. Especially for quantum computing
purposes \cite{Loss-98} coherent states are necessary to form and
couple qubits and to implement SWAP gates for qubit manipulation
\cite{Petta-05,Hanson-07}.

With dc-transport experiments there are only a few methods that
can give hints for molecular bonds. The width of anticrossings
visible in charging diagrams is a measure \cite{Golden-96},
although anticrossings appear for capacitively coupled dots as
well. In addition the curvature of the lines forming an
anticrossing can be used \cite{Blick-98} and also the visibility
of lines in non-parallel quantum dots \cite{Rogge-04}. Another
alternative is to study excited states. Strongly coupled quantum
dots form bonding and antibonding states that are visible in
nonlinear measurements \cite{Huttel-05}.

We have studied the impact of the coupling strength on the mean
charge of multiple quantum dot systems coupled in series. Using a
quantum point contact \cite{Field-93} we analyzed the mean charge
in stability diagrams. We found that depending on the symmetry of
tunneling rates and on the coupling strength characteristic
patterns are formed in nonlinear measurements. This allows to
non-invasively detect the symmetry of tunneling rates and the
quality of the interdot coupling.

\begin{figure}
 \includegraphics{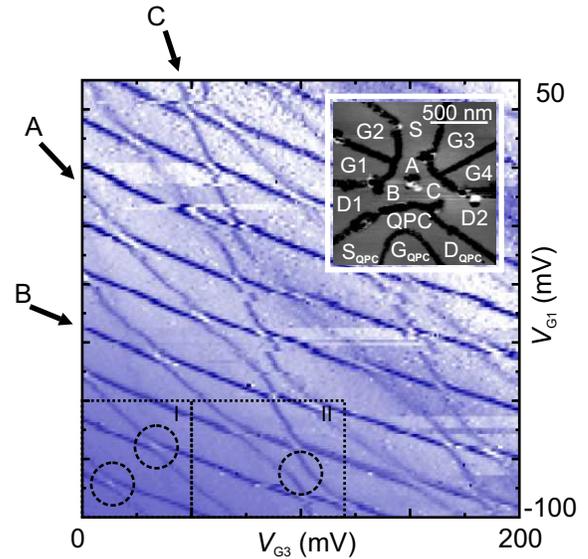}
 \caption{(color online) charging diagram of the triple dot device
 measured with charge detection. Three sets of lines appear
 from the three dots A, B and C. At the intersections anticrossings
 are visible due to interdot coupling (circles). Inset: atomic force
 microscopic image of the device with dots A, B, C
 and a quantum point contact for charge detection.}
 \label{fig1}
\end{figure}

The measurements were performed on a device containing three
quantum dots A, B and C (see inset of Fig. \ref{fig1}). The device
was produced using local anodic oxidation on a
GaAs/AlGaAs-heterostructure \cite{Ishii-95,Keyser-00}. The three
dots are positioned in a star-like geometry with one lead for each
dot (Source S at A, Drain1 D1 at B and Drain2 D2 at C). The
barriers and the potentials can be tuned with four sidegates G1 to
G4. A quantum point contact (QPC) is placed next to the three dots
for charge detection. The device is described in detail in Ref.
\cite{Rogge-08}.

Figure \ref{fig1} shows a charging diagram of the triple dot
device as a function of the voltages applied to gates G3 and G1.
The derivative of the QPC-current with respect to $V_{\text{G3}}$
is plotted. Dark lines correspond to an increase of charge
detected by the QPC (e.g. charging a dot with an electron), bright
features appear when the QPC detects a decrease of charge. Three
sets of lines are visible that denote charging of the three dots.
Lines with a shallow slope correspond to dot B, those with steap
slopes appear due to charging of dot C. Intermediate slopes
correspond to dot A. Whenever two lines from different sets
intersect, an anticrossing appears due to finite coupling between
the dots. At these anticrossings, two of the dots are in resonance
and can thus be treated as double quantum dot. From transport
measurement we know that the double dots A-B and A-C feature
finite interdot tunnel coupling, while the double dot B-C is
coupled capacitively only (see Ref. \cite{Rogge-08}). Therefore
B-C is not interesting for the purpose of this paper. In the
following we concentrate on the analysis of the two double dots
A-B and A-C. This analysis is done in the two sections I and II
with I showing anticrossings due to resonance of dots A and B, II
showing an anticrossing due to resonance of dots A and C
(circles).

\begin{figure}
 \includegraphics{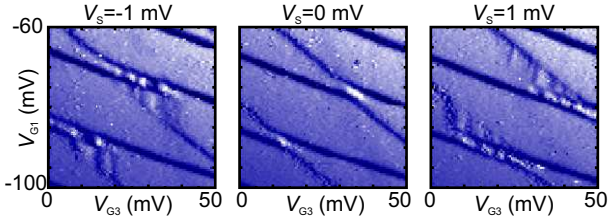}
 \caption{(color online) charge measurement in section I of Fig.
 \ref{fig1} for $V_{\text{S}}=-1$~mV, $V_{\text{S}}=0$~mV and $V_{\text{S}}=1$~mV. At
 $V_{\text{S}}\neq 0$ triangular patterns appear with additional lines of
 dark and bright features due to excited atomic states.}
 \label{fig2}
\end{figure}

Figure \ref{fig2} shows three graphs measured in the region of
section I. Charge detection is performed in the linear regime with
$V_{\text{S}}=0$ (center image) and in the nonlinear regime with
$V_{\text{S}}=-1$~mV (left) and $V_{\text{S}}=1$~mV (right). While
the center image shows the same pattern as observed in Fig.
\ref{fig1} with two sets of lines for dots A and B and two
anticrossings, the situation changes in the nonlinear regime
showing a more complex pattern with ground and excited states. The
anticrossings appear shifted to the upper left for
$V_{\text{S}}=-1$~mV and to the lower right for
$V_{\text{S}}=1$~mV. A triangular shaped pattern with additional
lines is connected to the right ($V_{\text{S}}=-1$~mV) and to the
left respectively ($V_{\text{S}}=1$~mV). These triangles are
familiar from nonlinear transport measurements in weakly coupled
quantum dots \cite{Dixon-96,Wiel-03} and have recently been
measured with charge detection as well \cite{Johnson-05}. However,
triangles with such patterns have not been reported so far. Dark
and bright features alternate corresponding to alternating
increase and decrease of mean charge measured with the QPC.

\begin{figure}
 \includegraphics{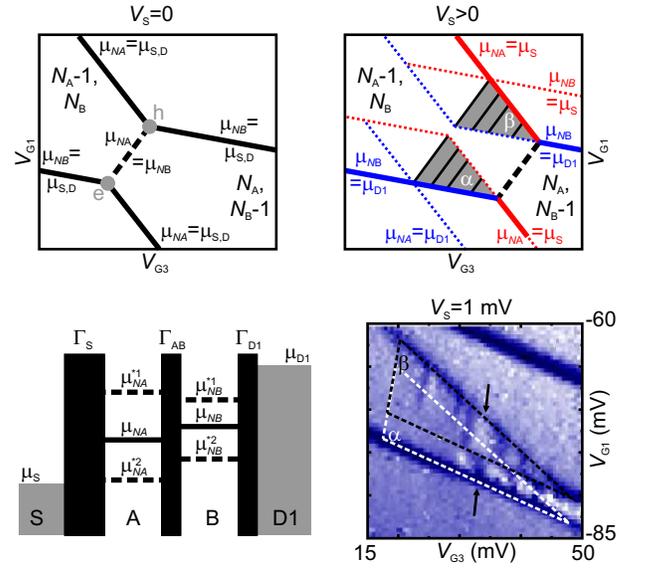}
 \caption{(color online) top left: schematic for the anticrossing
 of dots A and B for $V_{\text{S}}=0$ with triple points e and h. Top right: schematic for the
 anticrossing at $V_{\text{S}}>0$. Two triangles $\alpha$ and $\beta$
 appear with additional lines. Bottom left: possible configuration of chemical potentials. Bottom right: Section of Fig. \ref{fig2},
 $V_{\text{S}}=$1~mV. The two triangles $\alpha$ and $\beta$ are visible even
 though they overlap. The line pattern has the opposite order in
 both triangles. A dark line in $\alpha$ becomes bright in $\beta$.}
 \label{fig3}
\end{figure}

The origin of this pattern is explained with the schematics shown
in Fig. \ref{fig3}. Assuming the total number of electrons to be
$N_{\text{A}}-1$ or $N_{\text{A}}$ on dot A and $N_{\text{B}}-1$
or $N_{\text{B}}$ on dot B with ground state energies
$E_{N\text{A}-1}$ or $E_{N\text{A}}$ and $E_{N\text{B}-1}$ or
$E_{N\text{B}}$, two transitions are possible:
\begin{eqnarray*}
E_{N\text{A}-1}&\leftrightarrow& E_{N\text{A}}\text{ with chemical potential }\\
&&\mu_{N\text{A}}=E_{N\text{A}}-E_{N\text{A}-1},\\
E_{N\text{B}-1}&\leftrightarrow& E_{N\text{B}}\text{ with chemical potential }\\
&&\mu_{N\text{B}}=E_{N\text{B}}-E_{N\text{B}-1}.
\end{eqnarray*}
If these chemical potentials equal those of the leads
($\mu_{\text{S}}$ and $\mu_{\text{D1}}$) lines are visible in the
charging diagram. At $V_{\text{S}}=0$ (left schematic)
$\mu_{\text{S}}$ and $\mu_{\text{D1}}$ are degenerate. Therefore
each chemical dot potential produces a single line forming the
typical hexagonal cells with the so called triple points (marked
with e and h) at the edges. At e transport through the serial
double quantum dot can be described by sequential tunneling of one
electron at a time through the otherwise empty dot states. At h
transport occurs by sequential tunneling of one hole at a time
through the otherwise filled dot states. In between the two triple
points the chemical potentials of both dots are equal
($\mu_{N\text{A}}=\mu_{N\text{B}}$) and an electron can move from
dot B to dot A with increasing $V_{\text{G3}}$. As dot A is
further away from the QPC, the QPC detects a decrease of charge.
Thus a white feature is visible at the anticrossings in the center
image of Fig. \ref{fig2}.

The nonlinear regime is described with the schematic on the right
($V_{\text{S}}>0$). The discussion for $V_{\text{S}}<0$ is analog.
At $V_{\text{S}}>0$ the degeneracy of $\mu_{\text{S}}$ and
$\mu_{\text{D1}}$ is lifted, $\mu_{\text{D1}}>\mu_{\text{S}}$.
Therefore there are two possible resonance conditions for each
chemical dot potential. But as each dot does only couple to one
lead, only one resonance condition per dot is relevant. Thus still
only one line is visible per ground state transition (with
$\mu_{N\text{A}}=\mu_{\text{S}}$ and
$\mu_{N\text{B}}=\mu_{\text{D1}}$). The other resonance conditions
do not appear (dotted lines). However, as the two dots use
different chemical lead potentials the anticrossings are shifted
to the lower right as observed in the right image of Fig.
\ref{fig2}. Therefore the exchange of an electron between the dots
does not appear at $\mu_{N\text{A}}=\mu_{N\text{B}}$, but at the
dashed black line (right schematic). Left to this line there are
two triangles (grey) where both chemical potentials,
$\mu_{N\text{A}}$ and $\mu_{N\text{B}}$, are between
$\mu_{\text{S}}$ and $\mu_{\text{D1}}$ and
$\mu_{N\text{A}}<\mu_{N\text{B}}$ (Fig. \ref{fig3}, bottom left).
These are the triangles described above. At the left border of
these triangles the resonance condition
$\mu_{N\text{A}}=\mu_{N\text{B}}$ is fulfilled opening a transport
channel. Further transport channels within the triangles can
appear due to excited atomic states.

As an example we take into account two excited states per dot with
total energies $E_{N-1}^*>E_{N-1}$ and $E_{N}^*>E_{N}$. Now two
additional transitions are possible for each dot with new chemical
potentials (Fig. \ref{fig3}, bottom left):
\begin{eqnarray*}
E_{N-1}&\leftrightarrow& E_{N}^*\text{ with chemical potential }
\mu_{N}^{*1}>\mu_{N},\\
E_{N-1}^*&\leftrightarrow& E_{N}\text{ with chemical potential }
\mu_{N}^{*2}<\mu_{N}.
\end{eqnarray*}
Additional transport channels form for $V_{\text{S}}>0$, if the
following resonance conditions are fulfilled (other channels are
forbidden due to trapping):
\begin{eqnarray*}
\mu_{N\text{A}}&=&\mu_{N\text{B}}^{*2},\\
\mu_{N\text{A}}^{*1}&=&\mu_{N\text{B}},\\
\mu_{N\text{A}}^{*1}&=&\mu_{N\text{B}}^{*2}.
\end{eqnarray*}
Together with the resonance condition
$\mu_{N\text{A}}=\mu_{N\text{B}}$ those are the four solid lines
drawn in each grey triangle in the schematic. They can appear in
conductance measurements with electron-like transport in triangle
$\alpha$ and hole-like transport in triangle $\beta$.

With charge detection resonances are only visible, if they feature
a different mean charge than what is given in the grey regions.
Within the grey triangles at $V_{\text{S}}>0$ electrons can enter
dot B via Drain1, holes can enter dot A via Source (electrons can
leave A). Off resonance no transport between the dots is possible.
Therefore the mean charge in the grey regions, added to the charge
background of $N_{\text{A}}$ and $N_{\text{B}}$ electrons, is one
electron on dot B. On resonance, transport between the dots is
possible. Now the mean charge depends on the symmetry of the
tunneling rates $\Gamma_{\text{S}}$ between Source and dot A,
$\Gamma_{\text{D1}}$ between Drain1 and dot B and the interdot
tunneling rate $\Gamma_{\text{AB}}$. Three different symmetries
are possible that define the mean charge on resonance within
triangles $\alpha$ and
$\beta$:\\\\
(i) $\Gamma_{\text{AB}}<\Gamma_{\text{S}}$, $\Gamma_{\text{D1}}$:\\
\hspace*{0.5cm}$\alpha$: one electron on B.\\
\hspace*{0.5cm}$\beta$: one electron on B.\\
(ii) $\Gamma_{\text{D1}}<\Gamma_{\text{S}}$, $\Gamma_{\text{AB}}$:\\
\hspace*{0.5cm}$\alpha$: no electron on A and B.\\
\hspace*{0.5cm}$\beta$: one electron equally occupying both dots.\\
(iii) $\Gamma_{\text{S}}<\Gamma_{\text{AB}}$, $\Gamma_{\text{D1}}$:\\
\hspace*{0.5cm}$\alpha$: one electron equally occupying both dots.\\
\hspace*{0.5cm}$\beta$: one electron on both dots each.\\\\
In (i) no lines are visible with charge detection as the mean
charge on resonance is identical to the mean charge off resonance
in the grey regions. In (ii) and (iii) a resonance changes the
mean charge in both triangles and becomes visible. Thus it is much
more probable to observe excited states in weakly coupled double
dots than in single dots, where excited states can only appear
with symmetric tunneling rates \cite{Rogge-05}.

Therefore (ii) or (iii) must be true for the measurements
presented in Fig. \ref{fig2}. A more detailed analysis reveals the
actual ratio of tunneling rates. The bottom of Figure \ref{fig3}
shows a section of the right image in Fig. \ref{fig2} with the
triangles $\alpha$ and $\beta$ marked with white and black lines.
As the difference between $\mu_{\text{S}}$ and $\mu_{\text{D1}}$
is bigger than the splitting of the anticrossing, both triangles
overlap. Within the triangles the additional lines are visible.
Following the line marked with arrows from bottom to top, one
first observes a dark feature in triangle $\alpha$ and then a
bright feature in triangle $\beta$. Thus the effect on the mean
charge must be vice versa in both triangles. This is only possible
in (iii) with a decrease of mean charge in (iii)$\alpha$ (as dot B
is closer to the QPC than dot A) and an increase in (iii)$\beta$.

Within the overlapped region an electron can enter the possibly
empty dots via Drain1, as the system is within triangle $\alpha$.
This electron can now leave the dots again via Source, or another
electron can enter via Drain1, as the system is in triangle
$\beta$ as well. As $\Gamma_{\text{S}}<\Gamma_{\text{D1}}$, it is
much more probable for a second electron to enter the system than
for the first one to leave. Therefore the process related to
triangle $\beta$ is favored.

\begin{figure}
 \includegraphics{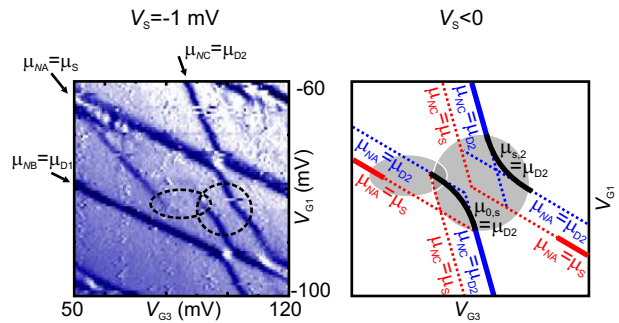}
 \caption{(color online) left: section II from Fig. \ref{fig1}
 at $V_{\text{S}}=-1$~mV. An anticrossing of dots A and C is visible
 (circle). Instead of triangular shaped patterns a
 splitting is observed on the left (ellipse). Right: schematic
 for the measurement on the left. The pattern is formed by a
 molecular state that is created at the anticrossing.}
 \label{fig4}
\end{figure}

Triangular patterns are visible for A-B over a wide range of
parameters. They finally fade out with increasing gate voltages as
the tunneling rates are changed. In contrast no such patterns
appear for A-C, although measured under the same conditions within
the same device. Instead a different pattern is found. The left of
Figure \ref{fig4} shows a measurement at $V_{\text{S}}=-1$~mV,
taken within section II (as the lines of dot A appear steeper than
those of B, but shallower than those of C, patterns of A-B at
$V_{\text{S}}>0$ must be compared with patterns of A-C with
$V_{\text{S}}<0$). The measurement shows an anticrossing (circle),
that is almost not shifted compared to the one observed at
$V_{\text{S}}=0$ (see Fig. \ref{fig1}). Another striking feature
is the step that appears on the left of the anticrossing
(ellipse). The left line of the anticrossing disappears and comes
up again with a huge offset to the left. There are no triangular
shaped patterns or lines for excited states.

The origin of this pattern is described using the schematic at the
right of Fig. \ref{fig4} assuming molecular bonds. With a relative
width of anticrossings of $\approx 0.4$ (with 1 being the maximum
\cite{Golden-96}) the double dot A-C is coupled stronger than A-B,
which has a relative width of ca. 0.33. As for the schematics
shown before for $V_{\text{S}}\neq 0$ the resonances for ground
state transitions split into two resonances as the chemical
potentials in the leads differ. Here with $V_{\text{S}}<0$
resonances with $\mu_{\text{S}}$ appear shifted to the lower left
compared to those with $\mu_{\text{D2}}$. Far off the anticrossing
states of the two dots can be described as atomic with chemical
potentials $\mu_{N\text{A}}$ and $\mu_{N\text{C}}$. As dot A is
coupled to Source and C to Drain2, the resonances
$\mu_{N\text{A}}=\mu_{\text{S}}$ and
$\mu_{N\text{C}}=\mu_{\text{D2}}$ must appear (solid straight
lines). Due to the strong coupling of A and C the pattern around
the anticrossing cannot be described with atomic states any
longer. Instead a common symmetric molecular state with energy
$E_{\text{s}}$ evolves that is extended over the whole double dot.
With the double dot energies $E_0$ for no added electrons and
$E_2$ for two electrons added, two new transitions are possible:
\begin{eqnarray*} E_0&\leftrightarrow& E_{\text{s}} \text{ with chemical
potential }
\mu_{\text{0,s}}<\mu_{N\text{A}}\text{, }\mu_{N\text{C}},\\
E_{\text{s}}&\leftrightarrow& E_2 \text{ with chemical potential }
\mu_{\text{s,2}}>\mu_{N\text{A}}\text{, }\mu_{N\text{C}}.
\end{eqnarray*}
These new chemical potentials can create two resonances each, one
with $\mu_{\text{S}}$, one with $\mu_{\text{D2}}$. Which of those
involves a change of the mean charge depends on the tunneling
rates again. If $\Gamma_{\text{D2}}<\Gamma_{\text{S}}$, charging
appears at resonance with $\mu_{\text{S}}$. If
$\Gamma_{\text{S}}<\Gamma_{\text{D2}}$, charging appears at
resonance with $\mu_{\text{D2}}$ instead. The latter case results
in the two curved solid lines in the schematic, that properly
describes the experiment. In the area close to the anticrossing
the double dot shows charging at resonance with $\mu_{\text{D2}}$
as well as dot C far off the anticrossing. Dot A shows charging at
resonance with $\mu_{\text{S}}$ instead. Therefore a jump must
occur when the system changes from the molecular common state to
the atomic state of dot A. Two of those jumps are shown in the
schematic, but only one is visible in the measurement as the other
one is disturbed by a line of dot B. However, the symmetry of
tunneling rates is detected for double dot A-C as well.

Thus with charge measurements it is possible to detect the
symmetry of tunneling rates for weakly and for strongly coupled
double quantum dots. For the two double dots in this device the
same symmetry was detected:
$\Gamma_{\text{S}}<\Gamma_{\text{AB,AC}}$,
$\Gamma_{\text{D1,D2}}$. However, depending on the strength of
tunnel coupling two completely different patterns were found. Thus
non-invasive charge measurement is capable of detecting molecular
bonds in quantum dots.

For the heterostructure we thank M. Bichler, G. Abstreiter, and W.
Wegscheider. This work has been supported by BMBF via nanoQUIT.




\begin{thebibliography}{20}
\expandafter\ifx\csname
natexlab\endcsname\relax\def\natexlab#1{#1}\fi
\expandafter\ifx\csname bibnamefont\endcsname\relax
  \def\bibnamefont#1{#1}\fi
\expandafter\ifx\csname bibfnamefont\endcsname\relax
  \def\bibfnamefont#1{#1}\fi
\expandafter\ifx\csname citenamefont\endcsname\relax
  \def\citenamefont#1{#1}\fi
\expandafter\ifx\csname url\endcsname\relax
  \def\url#1{\texttt{#1}}\fi
\expandafter\ifx\csname urlprefix\endcsname\relax\def\urlprefix{URL }\fi
\providecommand{\bibinfo}[2]{#2}
\providecommand{\eprint}[2][]{\url{#2}}


\bibitem[{\citenamefont{Kouwenhoven et~al.}(1997)\citenamefont{Kouwenhoven, Marcus, McEuen, Tarucha, Westervelt, and
Wingreen}}]{Kouwenhoven-97}
\bibinfo{author}{\bibfnamefont{L.~P.} \bibnamefont{Kouwenhoven}},
 \bibinfo{author}{\bibfnamefont{C.~M.} \bibnamefont{Marcus}},
 \bibinfo{author}{\bibfnamefont{P.~L.} \bibnamefont{McEuen}},
 \bibinfo{author}{\bibfnamefont{S.}~\bibnamefont{Tarucha}},
 \bibinfo{author}{\bibfnamefont{R.~M.} \bibnamefont{Westervelt}},
 \bibnamefont{and} \bibinfo{author}{\bibfnamefont{N.~S.} \bibnamefont{Wingreen}},
 in \emph{\bibinfo{booktitle}{Mesoscopic Electron
  Transport}}, edited by \bibinfo{editor}{\bibfnamefont{L.~L.}
  \bibnamefont{Sohn}}, \bibinfo{editor}{\bibfnamefont{L.~P.}
  \bibnamefont{Kouwenhoven}}, \bibnamefont{and}
  \bibinfo{editor}{\bibfnamefont{G.}~\bibnamefont{Sch\"o{}n}}
  (\bibinfo{publisher}{Kluwer}, \bibinfo{address}{Dordrecht},
  \bibinfo{year}{1997}), vol. \bibinfo{volume}{345} of
  \emph{\bibinfo{series}{Series E}}, pp. \bibinfo{pages}{105--214}.





\bibitem[{\citenamefont{van~der Wiel et~al.}(2003)\citenamefont{van~der Wiel,
  Franceschi, Elzerman, Fujisawa, Tarucha, and Kouwenhoven}}]{Wiel-03}
\bibinfo{author}{\bibfnamefont{W.~G.} \bibnamefont{van~der Wiel}},
 \bibinfo{author}{\bibfnamefont{S.~D.} \bibnamefont{Franceschi}},
 \bibinfo{author}{\bibfnamefont{J.~M.} \bibnamefont{Elzerman}},
 \bibinfo{author}{\bibfnamefont{T.}~\bibnamefont{Fujisawa}},
 \bibinfo{author}{\bibfnamefont{S.}~\bibnamefont{Tarucha}}, \bibnamefont{and}
 \bibinfo{author}{\bibfnamefont{L.~P.} \bibnamefont{Kouwenhoven}},
  \bibinfo{journal}{Rev. Mod. Phys.} \textbf{\bibinfo{volume}{75}},
  \bibinfo{pages}{1} (\bibinfo{year}{2003}).

  \bibitem[{\citenamefont{Blick et~al.}(1996)\citenamefont{Blick, Haug, Weis, Pfannkuche, v. Klitzing, and Eberl}}]{Blick-96}
\bibinfo{author}{\bibfnamefont{R.~H.} \bibnamefont{Blick}},
\bibinfo{author}{\bibfnamefont{R.~J.} \bibnamefont{Haug}},
\bibinfo{author}{\bibfnamefont{J.} \bibnamefont{Weis}},
\bibinfo{author}{\bibfnamefont{D.} \bibnamefont{Pfannkuche}},
\bibinfo{author}{\bibfnamefont{K.} \bibnamefont{v. Klitzing}},
  \bibnamefont{and} \bibinfo{author}{\bibfnamefont{K.}~\bibnamefont{Eberl}},
  \bibinfo{journal}{Phys. Rev. B} \textbf{\bibinfo{volume}{53}},
  \bibinfo{pages}{7899} (\bibinfo{year}{1996}).

  \bibitem[{\citenamefont{Loss and DiVincenzo}(1998)}]{Loss-98}
\bibinfo{author}{\bibfnamefont{D.}~\bibnamefont{Loss}} \bibnamefont{and}
  \bibinfo{author}{\bibfnamefont{D.~P.} \bibnamefont{DiVincenzo}},
  \bibinfo{journal}{Phys. Rev. A} \textbf{\bibinfo{volume}{57}},
  \bibinfo{pages}{120} (\bibinfo{year}{1998}).

    \bibitem[{\citenamefont{Petta et~al.}(2005)\citenamefont{Petta, Johnson, Taylor, Laird, Yacoby, Lukin, Marcus, Hanson, and Gossard}}]{Petta-05}
  \bibinfo{author}{\bibfnamefont{J.~R.} \bibnamefont{Petta}},
  \bibinfo{author}{\bibfnamefont{A.~C.} \bibnamefont{Johnson}},
    \bibinfo{author}{\bibfnamefont{J.~M.} \bibnamefont{Taylor}},
      \bibinfo{author}{\bibfnamefont{E.~A.} \bibnamefont{Laird}},
        \bibinfo{author}{\bibfnamefont{A.} \bibnamefont{Yacoby}},
          \bibinfo{author}{\bibfnamefont{M.~D.} \bibnamefont{Lukin}},
            \bibinfo{author}{\bibfnamefont{C.~M.} \bibnamefont{Marcus}},
  \bibinfo{author}{\bibfnamefont{M.~P.}~\bibnamefont{Hanson}},
\bibnamefont{and} \bibinfo{author}{\bibfnamefont{A.~C.} \bibnamefont{Gossard}},
  \bibinfo{journal}{Science} \textbf{\bibinfo{volume}{309}},
  \bibinfo{pages}{2180} (\bibinfo{year}{2005}).

    \bibitem[{\citenamefont{Hanson et~al.}(2007)\citenamefont{Hanson, Kouwenhoven, Petta, Tarucha, and Vandersypen}}]{Hanson-07}
\bibinfo{author}{\bibfnamefont{R.}~\bibnamefont{Hanson}},
 \bibinfo{author}{\bibfnamefont{L.~P.} \bibnamefont{Kouwenhoven}},
  \bibinfo{author}{\bibfnamefont{J.~R.} \bibnamefont{Petta}},
   \bibinfo{author}{\bibfnamefont{S.} \bibnamefont{Tarucha}},
\bibnamefont{and} \bibinfo{author}{\bibfnamefont{L.~M.~K.} \bibnamefont{Vandersypen}},
  \bibinfo{journal}{Rev. Mod. Phys.} \textbf{\bibinfo{volume}{79}},
  \bibinfo{pages}{1217} (\bibinfo{year}{2007}).

\bibitem[{\citenamefont{Golden and Halperin}(1996)}]{Golden-96}
\bibinfo{author}{\bibfnamefont{J.~M.} \bibnamefont{Golden}} \bibnamefont{and}
  \bibinfo{author}{\bibfnamefont{B.~I.} \bibnamefont{Halperin}},
  \bibinfo{journal}{Phys. Rev. B} \textbf{\bibinfo{volume}{54}},
  \bibinfo{pages}{16757} (\bibinfo{year}{1996}).

    \bibitem[{\citenamefont{Blick et~al.}(1998)\citenamefont{Blick, Pfannkuche, Haug, v. Klitzing, and Eberl}}]{Blick-98}
\bibinfo{author}{\bibfnamefont{R.~H.} \bibnamefont{Blick}},
\bibinfo{author}{\bibfnamefont{D.} \bibnamefont{Pfannkuche}},
\bibinfo{author}{\bibfnamefont{R.~J.} \bibnamefont{Haug}},
\bibinfo{author}{\bibfnamefont{K.} \bibnamefont{v. Klitzing}},
  \bibnamefont{and} \bibinfo{author}{\bibfnamefont{K.}~\bibnamefont{Eberl}},
  \bibinfo{journal}{Phys. Rev. Lett.} \textbf{\bibinfo{volume}{80}},
  \bibinfo{pages}{4032} (\bibinfo{year}{1998}).

 \bibitem[{\citenamefont{Rogge et~al.}(2004)\citenamefont{Rogge, Fuehner, Keyser and Haug}}]{Rogge-04}
\bibinfo{author}{\bibfnamefont{M.~C.} \bibnamefont{Rogge}},
\bibinfo{author}{\bibfnamefont{C.} \bibnamefont{F\"uhner}},
\bibinfo{author}{\bibfnamefont{U.~F.} \bibnamefont{Keyser}},
  \bibnamefont{and} \bibinfo{author}{\bibfnamefont{R.~J.}~\bibnamefont{Haug}},
  \bibinfo{journal}{Appl. Phys. Lett.} \textbf{\bibinfo{volume}{85}},
  \bibinfo{pages}{606} (\bibinfo{year}{2004}).

  \bibitem[{\citenamefont{H\"uttel et~al.}(2005)\citenamefont{Huttel, Ludwig, Lorenz, Eberl, and Kotthaus}}]{Huttel-05}
\bibinfo{author}{\bibfnamefont{A.~K.} \bibnamefont{H\"uttel}},
\bibinfo{author}{\bibfnamefont{S.}~\bibnamefont{Ludwig}},
\bibinfo{author}{\bibfnamefont{H.}~\bibnamefont{Lorenz}},
\bibinfo{author}{\bibfnamefont{K.}~\bibnamefont{Eberl}},
\bibnamefont{and} \bibinfo{author}{\bibfnamefont{J.~P.} \bibnamefont{Kotthaus}},
  \bibinfo{journal}{Phys. Rev. B} \textbf{\bibinfo{volume}{72}},
  \bibinfo{pages}{081310(R)} (\bibinfo{year}{2005}).

  \bibitem[{\citenamefont{Field et~al.}(1993)\citenamefont{Field, Smith, Pepper, Richie, Frost, Jones, and Hasko}}]{Field-93}
\bibinfo{author}{\bibfnamefont{M.} \bibnamefont{Field}},
\bibinfo{author}{\bibfnamefont{C.~G.}~\bibnamefont{Smith}},
\bibinfo{author}{\bibfnamefont{M.}~\bibnamefont{Pepper}},
\bibinfo{author}{\bibfnamefont{D.~A.}~\bibnamefont{Ritchie}},
\bibinfo{author}{\bibfnamefont{J.~E.~F.}~\bibnamefont{Frost}},
\bibinfo{author}{\bibfnamefont{G.~A.~C.}~\bibnamefont{Jones}},
\bibnamefont{and} \bibinfo{author}{\bibfnamefont{D.~G.} \bibnamefont{Hasko}},
  \bibinfo{journal}{Phys. Rev. Lett.} \textbf{\bibinfo{volume}{70}},
  \bibinfo{pages}{1311} (\bibinfo{year}{1993}).

\bibitem[{\citenamefont{Ishii and Matsumoto}(1995)}]{Ishii-95}
\bibinfo{author}{\bibfnamefont{M.}~\bibnamefont{Ishii}} \bibnamefont{and}
  \bibinfo{author}{\bibfnamefont{K.}~\bibnamefont{Matsumoto}},
  \bibinfo{journal}{Jpn. J. Appl. Phys.} \textbf{\bibinfo{volume}{34}},
  \bibinfo{pages}{1329} (\bibinfo{year}{1995}).

\bibitem[{\citenamefont{Keyser et~al.}(2000)\citenamefont{Keyser, Schumacher,
  Zeitler, Haug, and Eberl}}]{Keyser-00}
\bibinfo{author}{\bibfnamefont{U.~F.} \bibnamefont{Keyser}},
 \bibinfo{author}{\bibfnamefont{H.~W.} \bibnamefont{Schumacher}},
 \bibinfo{author}{\bibfnamefont{U.}~\bibnamefont{Zeitler}},
 \bibinfo{author}{\bibfnamefont{R.~J.} \bibnamefont{Haug}}, \bibnamefont{and}
 \bibinfo{author}{\bibfnamefont{K.}~\bibnamefont{Eberl}},
  \bibinfo{journal}{Appl. Phys. Lett.} \textbf{\bibinfo{volume}{76}},
  \bibinfo{pages}{457} (\bibinfo{year}{2000}).

  \bibitem[{\citenamefont{Rogge et~al.}(2008)\citenamefont{Rogge and Haug}}]{Rogge-08}
\bibinfo{author}{\bibfnamefont{M.~C.} \bibnamefont{Rogge}}
  \bibnamefont{and} \bibinfo{author}{\bibfnamefont{R.~J.}~\bibnamefont{Haug}},
  \bibinfo{journal}{Phys. Rev. B} \textbf{\bibinfo{volume}{77}},
  \bibinfo{pages}{193306} (\bibinfo{year}{2008}).

  \bibitem[{\citenamefont{Dixon et~al.}(1996)\citenamefont{Dixon, Kouwenhoven, McEuen, Nagamune, Motohisa, Sakaki}}]{Dixon-96}
\bibinfo{author}{\bibfnamefont{D.} \bibnamefont{Dixon}},
\bibinfo{author}{\bibfnamefont{L.~P.}~\bibnamefont{Kouwenhoven}},
\bibinfo{author}{\bibfnamefont{P.~L.}~\bibnamefont{McEuen}},
\bibinfo{author}{\bibfnamefont{Y.}~\bibnamefont{Nagamune}},
\bibinfo{author}{\bibfnamefont{J.}~\bibnamefont{Motohisa}},
\bibnamefont{and} \bibinfo{author}{\bibfnamefont{H.} \bibnamefont{Sakaki}},
  \bibinfo{journal}{Phys. Rev. B} \textbf{\bibinfo{volume}{53}},
  \bibinfo{pages}{12625} (\bibinfo{year}{1996}).

   \bibitem[{\citenamefont{Johnson et~al.}(2005)\citenamefont{Johnson, Petta, Marcus, Hanson, and Gossard}}]{Johnson-05}
\bibinfo{author}{\bibfnamefont{A.~C.} \bibnamefont{Johnson}},
\bibinfo{author}{\bibfnamefont{J.~R.}~\bibnamefont{Petta}},
\bibinfo{author}{\bibfnamefont{C.~M.}~\bibnamefont{Marcus}},
\bibinfo{author}{\bibfnamefont{M.~P.} \bibnamefont{Hanson}},
  \bibnamefont{and} \bibinfo{author}{\bibfnamefont{A.~C.}~\bibnamefont{Gossard}},
  \bibinfo{journal}{Phys. Rev. B} \textbf{\bibinfo{volume}{72}},
  \bibinfo{pages}{165308} (\bibinfo{year}{2005}).

  \bibitem[{\citenamefont{Rogge et~al.}(2005)\citenamefont{Rogge, Harke, Fricke, Hohls, Reinwald, Wegscheider, and Haug}}]{Rogge-05}
\bibinfo{author}{\bibfnamefont{M.~C.} \bibnamefont{Rogge}},
\bibinfo{author}{\bibfnamefont{B.}~\bibnamefont{Harke}},
\bibinfo{author}{\bibfnamefont{C.}~\bibnamefont{Fricke}},
\bibinfo{author}{\bibfnamefont{F.} \bibnamefont{Hohls}},
\bibinfo{author}{\bibfnamefont{M.}~\bibnamefont{Reinwald}},
  \bibinfo{author}{\bibfnamefont{W.}~\bibnamefont{Wegscheider}},
  \bibnamefont{and} \bibinfo{author}{\bibfnamefont{R.~J.}~\bibnamefont{Haug}},
  \bibinfo{journal}{Phys. Rev. B} \textbf{\bibinfo{volume}{72}},
  \bibinfo{pages}{233402} (\bibinfo{year}{2005}).

\end{thebibliography}



\end{document}